\documentclass[epj]{svjour}
\usepackage{graphics}
\begin{document}
\title{Light Vector Mesons}
\author{Alexander Milov\inst{1}}
%}                     % Do not remove
%
\offprints{}          % Insert a name or remove this line
\institute{Weizmann Institute of Science, Rehovot, Israel, 76100}
\date{Received: date / Revised version: date}
\abstract{
This article reviews the current status of experimental results obtained in the measurement of light vector mesons produced in proton-proton and heavy ion collisions at different energies. The review is focused on two phenomena related to the light vector mesons; the modification of the spectral shape in search of chiral symmetry restoration and suppression of the meson production in heavy ion collisions. The experimental results show that the spectral shape of light vector mesons are modified compared to the parameters measured in vacuum. The nature and the magnitude of the modification depends on the energy density of the media in which they are produced. The suppression patterns of light vector mesons are different from the measurements of other mesons and baryons. The mechanisms responsible for the suppression of the mesons are not yet understood. Systematic comparison of existing experimental results points to the missing data which may help to resolve the problem.
\PACS{
      {PACS-key}{describing text of that key}   \and
      {PACS-key}{describing text of that key}
     } % end of PACS codes
} %end of abstract
\maketitle
\section{Introduction}
\label{intro}
Interest in the measurements of the properties of Low Mass Vector mesons (LVM) in Heavy Ion Collisions (HIC) has several reasons. First of all the bulk production of the LVM contributes approximately 15\% of the total particle production at the highest energies presently available at RHIC. Precise knowledge of the LVM production rates and spectral functions are therefore a mandatory condition to understand the nature of HIC. Specific properties of several LVM such as short lifetime and di-lepton decay modes singles them as an entirely unique class of probes to study different time slices of the hot and dense media created in HIC. The possibility to compare the di-lepton decay channels to the hadron decay channels allows understanding of physical processes dominating at different stages of the hot and dense medium evolution. Short life time and low interaction cross section of the lepton decay products allows testing the hypothesis of chiral symmetry restoration possibly taking place during the strongly coupled Quark Gluon Plasma (sQGP) stage of collision. A change in the spectral shape or in the branching ratio of the LVM may give evidence that chiral symmetry is restored in the interactions of relativistic Heavy Ions or is partially present in Cold Nuclear Matter.

Suppression of the high $p_{T}$ yield in relativistic HIC appears to be one of the most fascinating phenomena observed at RHIC in early runs and one of the major arguments that lead to the discovery of the sQGP. This phenomenon was further studied in great details by all RHIC experiments. It appears that at transverse momentum above 5~GeV/$c$ the magnitude of the suppression based on the existing data converges to some universal constant of 0.2-0.3. At the same time, below 5~GeV/$c$ most particles exhibit different behavior. The suppression patterns are not yet fully understood. Study of the LVM provides important input which allows understanding the suppression mechanisms by comparing particles with different masses, lifetimes and quark content. A systematic investigation of the particle production in the intermediate $p_{T}$ region requires high quality information about the production rates of LVM in different collision systems. This information becomes now available from PHENIX and STAR experiments studying the HIC at RHIC.

\section{Invariant mass spectra}
\label{sec:1}
This section compares and discusses the results of several experiments studying the LVM in the nuclear medium. The experimental groups measure the invariant mass spectra and the production rate of the LVM decaying in leptonic and hadronic channels. Different experiments use various nuclear targets and projectiles with different incident energies to initiate the reaction. For systematization purpose all experiments can be categorized in three groups according to the energy density which can be achieved in the colliding systems at their facilities. The experiments in three group study the cold nuclear matter, the intermediate energy region (SPS) and the highest energy density regime achievable RHIC. The energy density in the collisions can be approximately estimated based on the mass and energy of the colliding species by implying commonly used Bjorken formalism. 

\subsection{Cold nuclear matter measurements}
\label{sec:1.1}
\begin{figure*}[ht]
\resizebox{0.99\textwidth}{!}{
\includegraphics{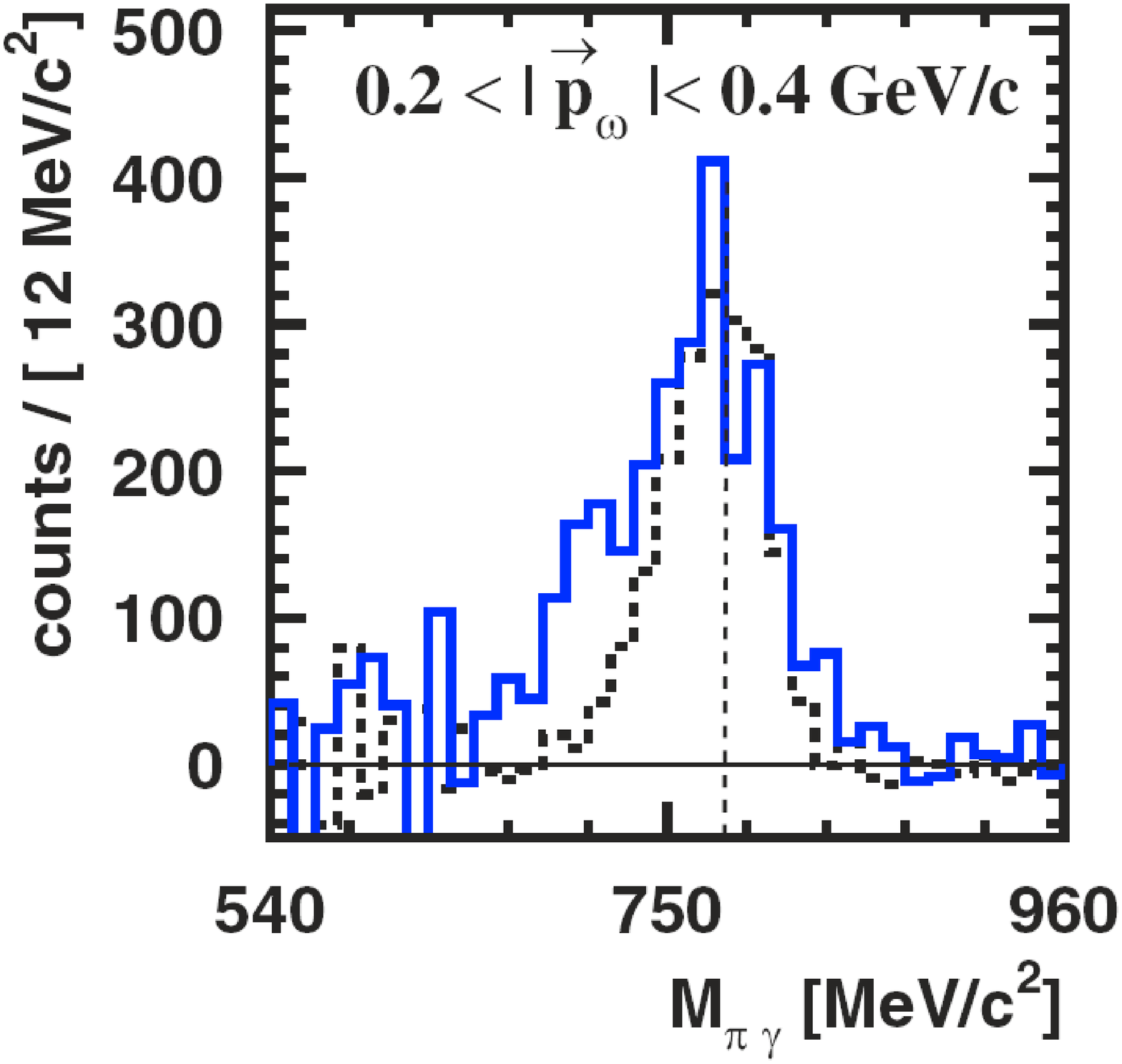}\\
\includegraphics{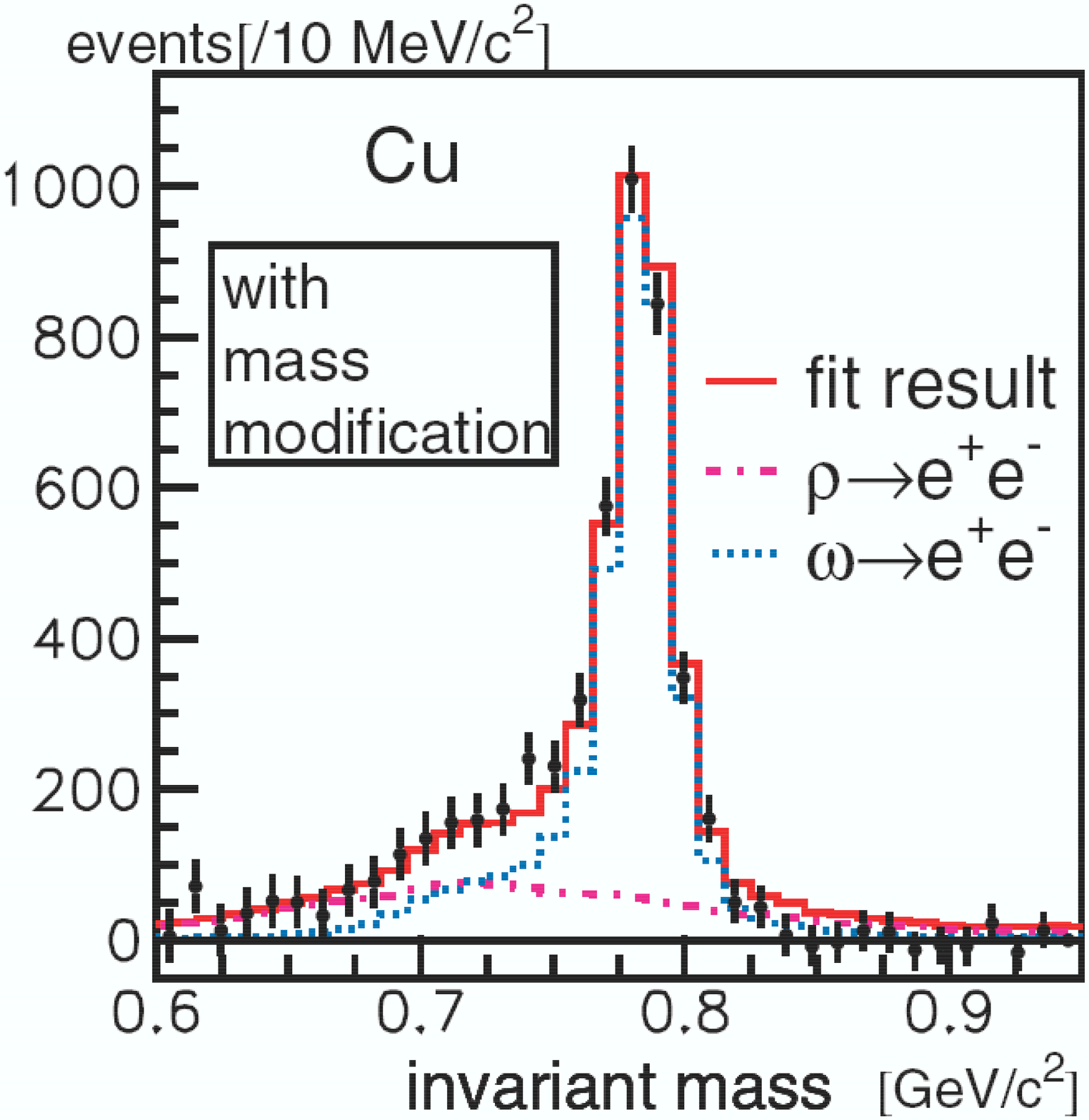}\\
\includegraphics{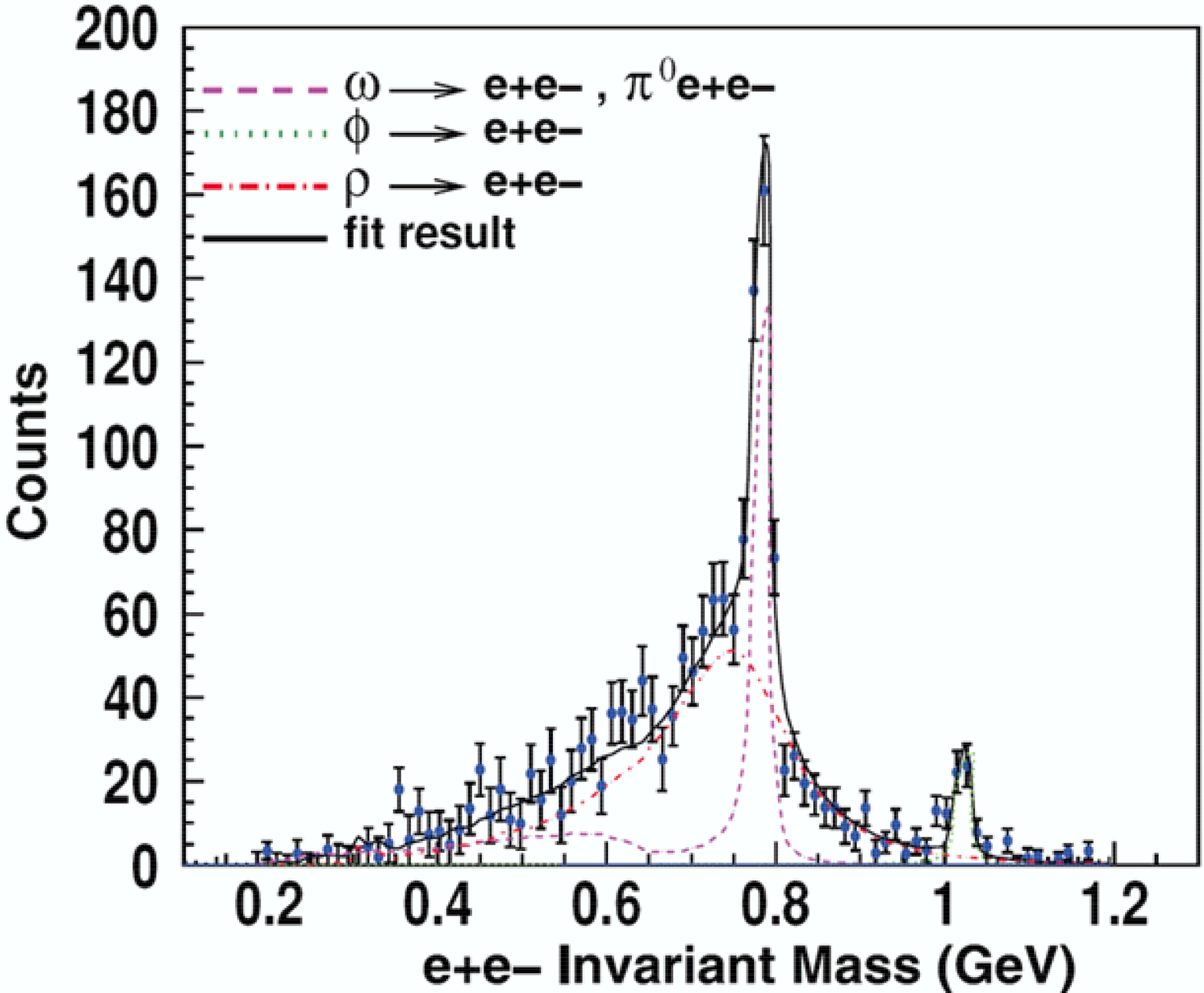}
}
\caption{Shown in the left panel are the CB-TAPS results of the $\pi^{0}\gamma$ invariant mass spectrum measurement after background subtraction and final state interaction suppression. The data are shown for the lowest transverse momentum bin of the $\omega$-meson emitted from $^{93}$Nb (solid) and LH$_{2}$ (dashed) targets~\cite{cb-taps}. The middle panel presents the results of the KEK E325 experiment measurement of the invariant $e^{+}e^{-}$ mass spectra after subtraction of the combinatorial background and removal of the $\eta$ and $\omega$ Dalitz decays. The data are shown together with model calculations considering the in-medium modification of the resonances~\cite{kek-ps1}. The right panel shows measured invariant $e^{+}e^{-}$ mass spectra published the CLAS collaboration using an electron beam incident on a $^{48}$Ti and $^{56}$Fe target. The combinatorial background is also removed.
}
\label{fig:cnm}
\end{figure*}

Four experiments in the first group study modification of the LVM invariant mass shape under experimental conditions in which no significant increase of the nuclear density can be achieved. The energy density in the media remains on the level of the cold nuclear matter density, approximately equal to 0.15~GeV/fm$^{3}$. The CB-TAPS experiment~\cite{cb-taps} measures the $\omega\rightarrow\pi^{0}\gamma$ decay channel in reactions induced by a photon incident upon various targets. The experiment observes a pronounced in-medium modification of the $\omega$-meson mass shape with the particles emitted from the $^{93}$Nb target. One of the CB-TAPS results is shown in the left panel of Fig.~\ref{fig:cnm}. The $\omega$-meson mass shape is different from the shape measured in photo-reactions on a lighter target such as LH$_{2}$. The mass of the $\omega$-meson measured in this reaction is also shown in the figure with the dashed line, showing no spectral shape modification. Several control measurement performed with long living mesons such as $\pi^{0}$, $\eta$ and $\eta\prime$ do not show modification of the shapes of the mesons beyond the apparatus effects. The enhancement seen in the lower part of the resonance peak gradually disappears with increasing transverse momentum of the $\omega$-meson.

The E325 experiment at KEK-PS measures the spectral shape of $\rho$-, $\omega$- and $\phi$-mesons~\cite{kek-ps1,kek-ps2} in 12~GeV/$c$ proton induced reactions on $^{64}$Cu and $^{12}$C targets. Unlike the CB-TAPS experiment, the E325 analyzes the di-electron channel which is less affected by final state interactions. The $\rho,\omega \rightarrow e^{+}e^{-}$ invariant mass spectra are shown in the middle panel of Fig.~\ref{fig:cnm}. The E325 experiment observes a modification of the invariant mass peaks of all three vector mesons in the lower part of the particle momentum distributions. The effect is observed in the reactions with light $^{12}$C and heavy $^{64}$Cu targets. For the $\rho$- and $\omega$-mesons the modification of the meson masses observed in this reaction is consistent for both targets. For the $\phi$-meson the modification of the mass in the $^{12}$C reaction is smaller. The E325 results support the scenario in which the mass of the particles drops proportionally to the nuclear density of the media.

The CLAS experiment at JLab recently published results of $\rho$-, $\omega$- and $\phi$-meson~\cite{clas} measurements in reactions induced by the 3-4~GeV electron beam of the CEBAF complex incident on a variety of targets ranging from $^{2}$H up to $^{48}$Ti-$^{56}$Fe. As in the E325 set-up, the CLAS detector measured LVM via the $e^{+}e^{-}$ decay channel. The $\rho$-meson peak measured by the CLAS collaboration after removal of the combinatorial background and $\omega$- $\phi$-meson peaks is shown in the right panel of Fig.~\ref{fig:cnm}. 

In spite of evident similarities between the experimental data shown in the middle and right panels of Fig.~\ref{fig:cnm} the conclusions derived by the CLAS collaboration differ from those claimed by the E325 collaboration. As a result of their analysis approach the CLAS collaboration observes no dropping of the $\rho$-mass or the masses of other particles with momentum above 0.8~GeV/$c$. Extraction of the spectral functions is under way. A slight broadening of the $\rho$-meson mass seen by the CLAS experiment is assigned to collision broadening and there are no signs of further modifications.

The TAGX collaboration at the Tokyo Electron Synchrotron measured sub-threshold $\rho$-meson production in the $\pi^{+}\pi^{-}$ decay channel induced by photons of 0.6-1.12~GeV on $^{2}$H, $^{3}$He and $^{12}$C targets. The choice of low mass targets allowed minimizing the effect of final state interactions. Based on the analysis of their data the TAGX collaboration finds the results to be consistent with a reduction of the $\rho$-meson invariant mass and providing no indication of the broadening of the $\rho$-meson peak.

\subsection{Intermediate energies}
\label{sec:1.2}
\begin{figure}[h]
\resizebox{0.49\textwidth}{!}{
\includegraphics{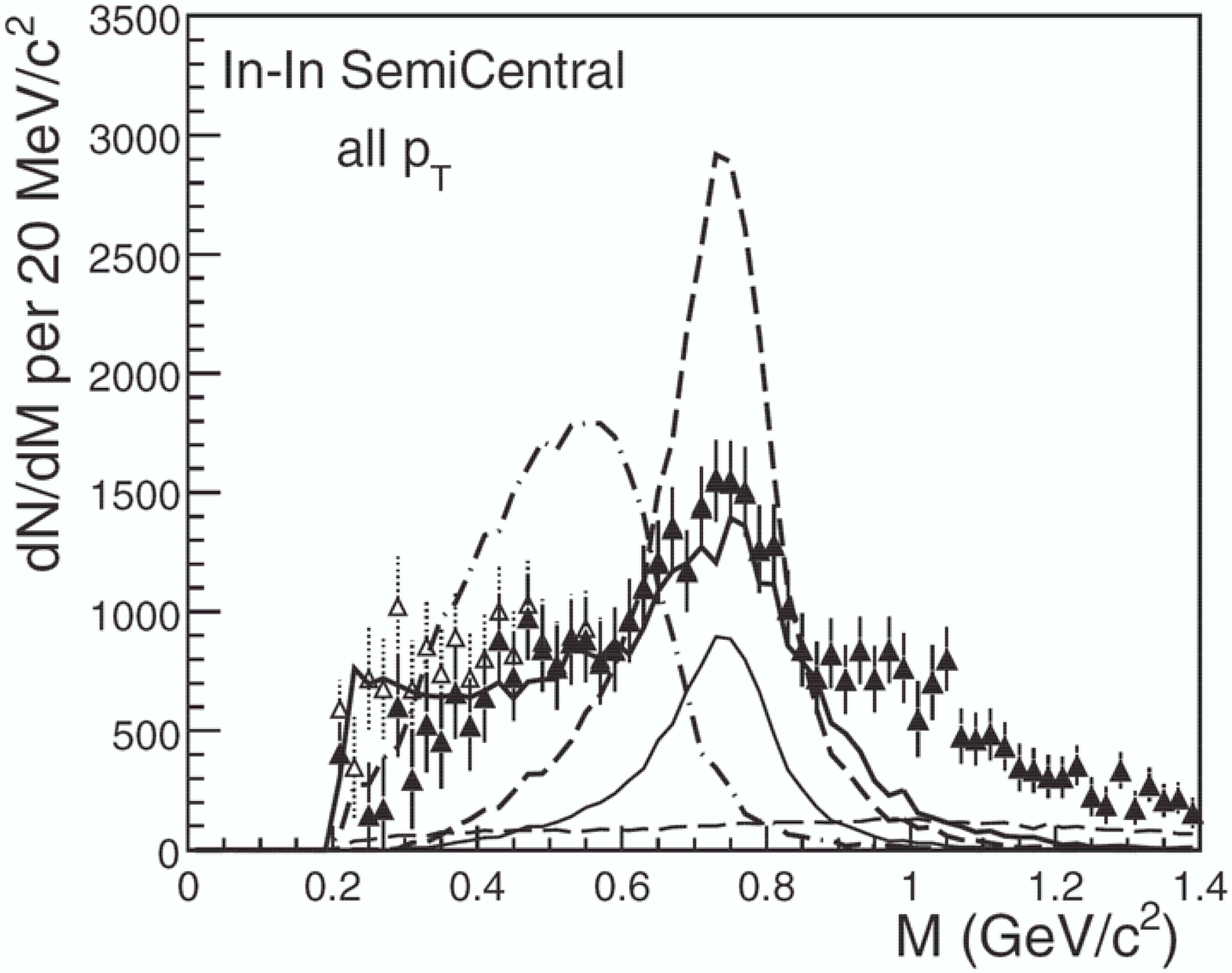}\\
\includegraphics{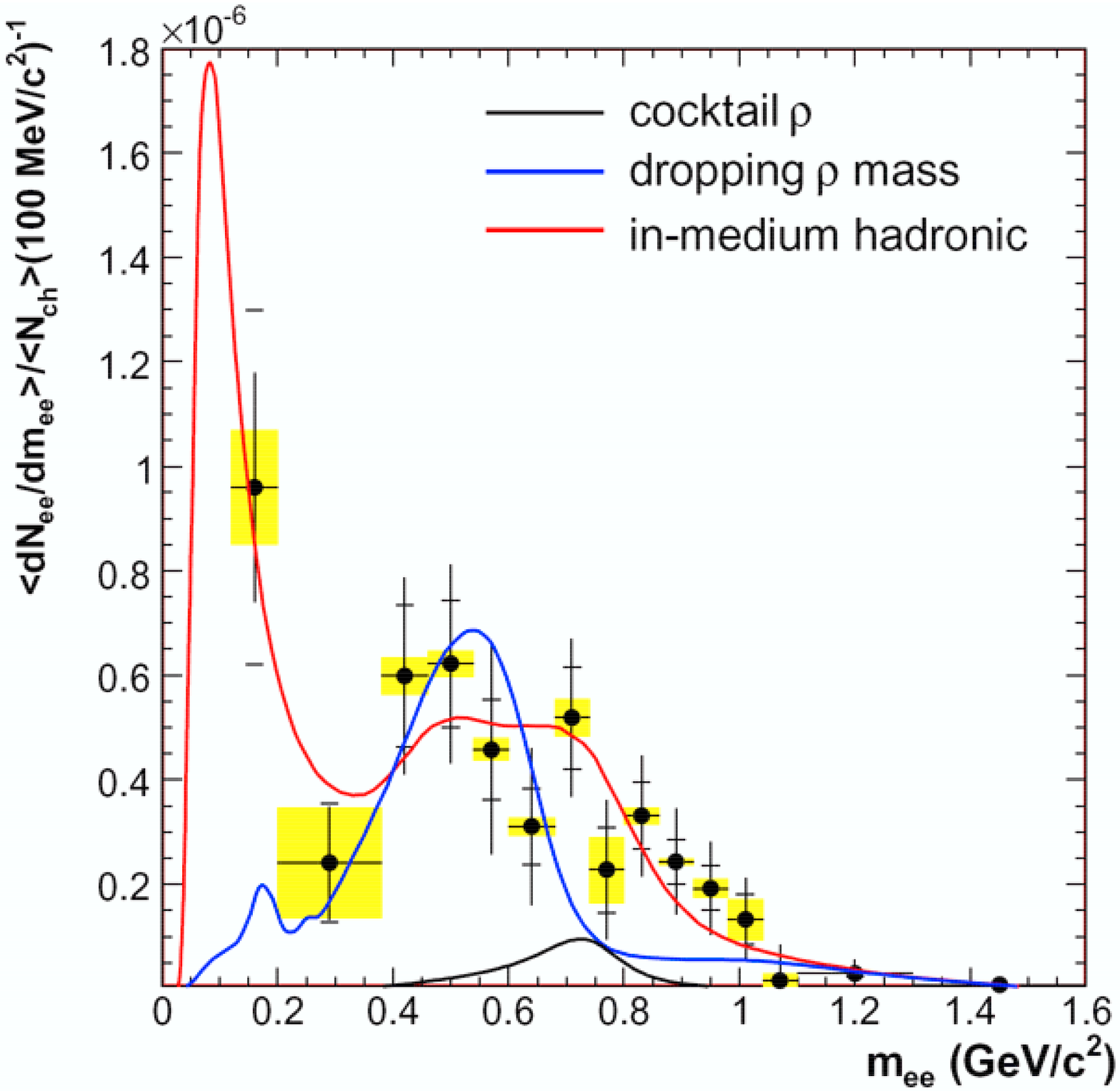}
}
\caption{Left panel shows the NA60 $\mu^{+}\mu^{-}$ invariant mass spectrum measured in $^{115}$In+$^{115}$In collisions at full SPS energy in the event multiplicity class $dN_{ch}/d\eta$=140. Combinatorial background and all resonances, except $\rho$ have been subtracted. The results are compared to the Cocktail $\rho$ (thin solid line), unmodified $\rho$ (dashed line), in-medium broadening $\rho$ (thick solid line), in-medium moving $\rho$ (dashed-dotted line). The errors are purely statistical. The systematic errors of the continuum are about 25\%. The figure is taken from~\cite{na60-1}. Right panel shows the $e^{+}e^{-}$ invariant mass yield after subtraction of the hadronic cocktail (without the $\rho$) measured by the CERES experiment. The systematic errors of the data are shown with the horizontal ticks and the systematic uncertainty of the subtracted cocktail with shaded boxes. The data are compared to the model predictions assuming shifting of the in-medium meson mass (blue online) and peak broadening scenario (red online). The figure is taken from~\cite{ceres-4}.}
\label{fig:ie}
\end{figure}

The two experiments dedicated to the measurement of low mass di-leptons at SPS are the CERES/NA45 and the NA60 collaborations. The NA60 recently published results on the $\rho$-meson in $\mu^{+}\mu^{-}$ decay channel in semi-central $^{115}$In+$^{115}$In collisions at full SPS energy~\cite{na60-1} of 158A~GeV. The invariant mass distribution in the $\rho$-meson peak region is shown in the left panel of Fig.~\ref{fig:ie}. The combinatorial background and $\omega$-, $\phi$- and $\eta$-meson contributions are removed. The experimental results are compared with different theoretical scenarios shown by lines\cite{rrapp}. From these comparisons it becomes clear that the shape of the $\rho$-meson peak is different from that measured in vacuum. The $\rho$-meson production rate increases with centrality faster compared to the rates of other particles. Unlike in cold nuclear matter, at SPS energies the dropping of the $\rho$-meson mass is not supported by the data. In the control measurement performed by NA60 with p+A collisions the $\mu^{+}\mu^{-}$ spectrum is well described by the decays of the resonances without modification of the their properties~\cite{na60-3}. Preliminary results reported by the NA60 collaboration point out that the $\omega$-meson production below $p_{T}$=0.5~GeV/$c$ can be suppressed~\cite{na60-4}.

The NA45/CERES experiment measured $e^{+}e^{-}$ invariant mass spectrum in the low mass region in a variety of collision systems at different SPS energies. No modification of the $e^{+}e^{-}$ mass spectra beyond the superposition of known sources was registered in p+A collisions\cite{ceres-1}. In $^{207}$Pb+$^{207}$Pb($^{197}$Au) collisions the CERES experiment observed enhancement in the low mass region both at full SPS energy of 158A~GeV~\cite{ceres-2} and at 40A~GeV~\cite{ceres-3}. Statistical significance of the CERES data is not that high as of the NA60, however recent reanalysis of the CERES data~\cite{ceres-4} shows that the enhancement in the di-electron channel is consistent with the measurement of the NA60 experiment. This result is shown in the right panel of Fig.~\ref{fig:ie}. One can see that the measurement is also consistent with the scenario of the broadening of the $\rho$-meson peak and does not support dropping of the $\rho$-mass. Modification of the invariant mass spectra depends on centrality and on the transverse momentum of the pair. The strongest effect in 158A~GeV collisions is seen for the pair $p_{T}$ below 0.5~GeV/$c$ and in the most central bin with the charged particle multiplicity of $dN_{ch}/d\eta$=370.

\subsection{High energy region}
\label{sec:1.3}

The PHENIX and STAR experiments at RHIC measured the LVM production in hadronic and di-electron decay modes. A number of measurements with hadronic decays were carried out under conditions where no modification of the invariant mass spectra was anticipated. Position and width of the $\phi\rightarrow K^{+}K^{-}$ peak in $^{197}$Au+$^{197}$Au collisions at 200~GeV in all measured centrality bins~\cite{phenix-1} are in agreement with the PDG values. Position of the $\omega$-meson peak in 200~GeV p+p and d+Au collisions~\cite{phenix-2} above $p_{T}$ of 2.5~GeV/$c$ was found unmodified. The STAR experiment published results for $\rho\rightarrow\pi^{+}\pi^{-}$~\cite{star-1} and $K^{*}\rightarrow K\pi$~\cite{star-2} measurements registering a dropping of the particle mass of these two resonances. The effect is observed in HIC collisions and in p+p collisions at $\sqrt{s_{NN}}$=62~and~200~GeV. Dropping of the particle masses decreases with the increase of the measured $p_{T}$ and vanishes at $p_{T}$=~1.5~GeV/$c$. The magnitude of the effect at $p_{T}$ approaching zero exceeds 40 and 10~MeV/$c^{2}$ for $\rho$ and $K^{*}$, respectively, which is approximately twice larger than the systematic uncertainly of the measurement.

\begin{figure}
\resizebox{0.49\textwidth}{!}{
\includegraphics{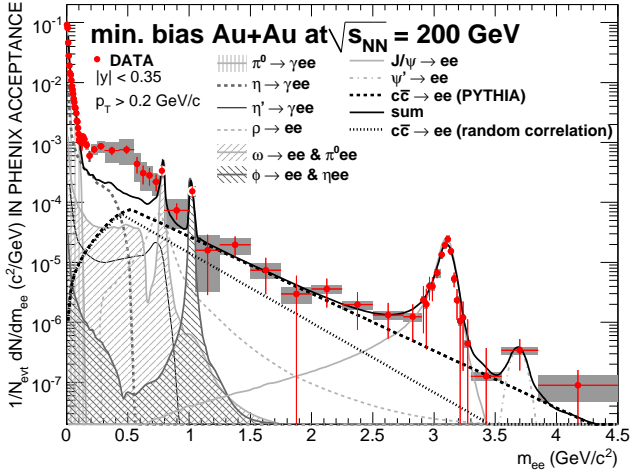}
}
\caption{Measured invariant mass yield of $e^{+}e^{-}$ pair compared to the superposition of know sources from hadron decays, the cocktail. The charmed meson decay contribution is calculated based on PYTHIA generator. Statistical (bars) and systematic (boxes) uncertainties are shown separately. The mass range covered by each data point is given by horizontal bars. The systematic uncertainty on the cocktail is not shown. The plot is published in~\cite{phenix-4}.}
\label{fig:he}
\end{figure}

The PHENIX experiment measured the di-electron invariant mass spectrum in p+p~\cite{phenix-3} and in $^{197}$Au+$^{197}$Au~\cite{phenix-4} collisions at full RHIC energy. The result of the Au+Au measurement is shown in Fig.~\ref{fig:he}. The PHENIX collaboration observes a significant enhancement in the low mass region in Au+Au collisions which increases with centrality. No such effect is registered in the p+p interactions at the same energy. Although the enhancement is similar to that observed at the SPS by the CERES and NA60 experiments (Fig.~\ref{fig:ie}) all theoretical models which successfully explain the phenomena at SPS fail to reproduce the enhancement measured at RHIC. See publication~\cite{phenix-4} for references.

To summarize, there is strong experimental evidence that the masses of the LVM produced inside nuclear matter are modified. The signature and the magnitude of the observed modification depend on the energy density. Table~\ref{tab:1} presents a list of the main results from different experiments.
\begin{table}
\caption{Summary of the LVM modification measurements performed by different experiments. The energy density in the media ($\rho$), if not provided by the experiment, is estimated using Bjorken formalism. Analyzed LVM and momentum range of the observed effect ($p$, or $p_{T})$ are listed in the third and fourth columns. Notations of ``D'', ``B'' and ``E'' in the 5th column are used to denote the effect of ``Dropping of the mass'', ``Broadening of the width'' and ``Enhancement of the production rate'' respectively.}
\label{tab:1}  
\begin{tabular}{llllll}
\hline\noalign{\smallskip}
Experi- &       $\rho$         & LVM of              & $p,p_{T}$         & Main   & Ref. \\
ment    & $\frac{GeV}{fm^{3}}$ & study               & $\frac{GeV}{c}$   & effect &      \\
\noalign{\smallskip}\hline\noalign{\smallskip}			       									 
CB-TAPS    & $\sim$0.1         & $\omega$            & $\le$0.5          & D      & \cite{cb-taps} \\
E325       & $\sim$0.15        & $\rho,\omega,\phi$  & $\le$1.0          & D, noB & \cite{kek-ps1,kek-ps2} \\
CLAS       & $\sim$0.15        & $\rho$              & $\ge$0.8          & noD    & \cite{clas} \\
TAGX       & $\sim$0.15        & $\rho$              & $\le$0.35         & D, noB & \cite{tagx} \\
NA60       & $\le$1.5          & $\rho$            & $\le$2.0          & B \& E & \cite{na60-1,na60-2} \\
CERES      & $\sim$1-2         & $\rho$              & $\le$0.5          & B \& E & \cite{ceres-4} \\
STAR       & $\sim$1-5         & $\rho$, $K^{*}$     & $\le$1.5          & D      & \cite{star-1,star-2} \\
PHENIX     & $\sim$1-5         & below $\rho$        & n/a               & E      & \cite{phenix-4} \\
\noalign{\smallskip}\hline
\end{tabular}
% Or use
\vspace*{5cm}  % with the correct table height
\end{table}
 In cold nuclear matter the main effect measured by several experimental groups favors dropping of the resonance masses. At intermediate SPS energies the production of the $\rho$-meson is significantly enhanced and the width of its peak is broader than in the vacuum. At the same time the data does not support shifting of the $\rho$-mass. At RHIC energies measurements of short-living particles ($\rho, K^{*}$) in hadron decay channels show sign of dropping mass effects in all collisions systems at low $p_{T}$, whereas measurement of $\phi$ or $\omega$ at higher $p_{T}$ yields unmodified values. The di-electron continuum measured at RHIC reveals a significant enhancement in the mass region around 0.4~GeV/$c^{2}$. The results of different experiments measured under similar energy density are usually consistent with each other. At the moment there is no single model capable to explain the phenomena measured under different energy density conditions.

\section{Spectra and ratios}
\label{sec:2}

New results on the spectral measurements of the LVM are given in this section. Because of the abundance of the data only selected spectra from the RHIC experiments are presented.

\subsection{Invariant yields}
\begin{figure}
\resizebox{0.49\textwidth}{!}{
\includegraphics{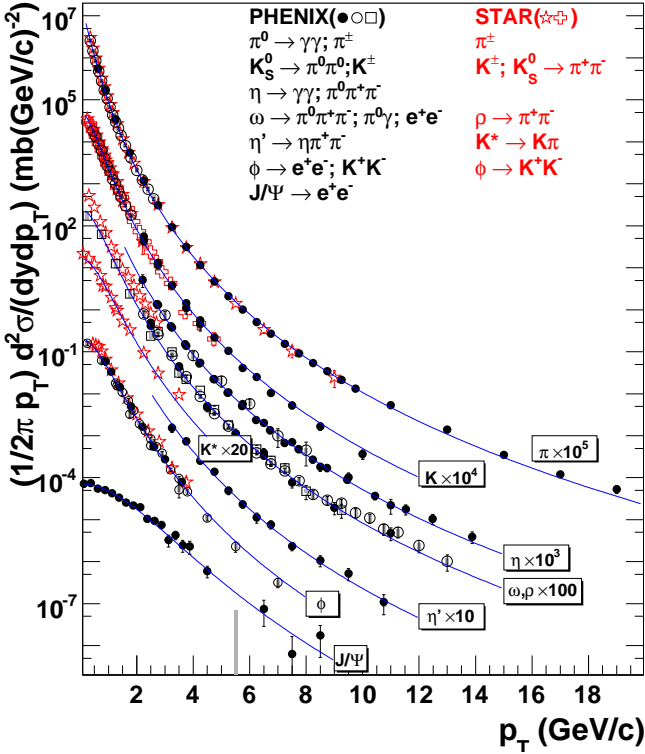}}
\caption{Compilation of invariant $p_{T}$ cross-section spectra measured in p+p collisions at $\sqrt{s}$=200~GeV at RHIC. The published and preliminary PHENIX data top-to-bottom are for $\pi$, K, $\eta$, $\omega$, $\eta\prime$, $\phi$ and J/$\Psi$ (black online). The data ware measured using Run3-Run5 event samples and in different decay channels. The data are taken from references~\cite{ppg30,ppg55,ppg63,ppg64,ppg69,vic,max,nakamiya}. The published STAR data shown top-to-bottom for $\pi$, K, $\rho$, K$^{*}$ and $\phi$ (red online) are extracted from references~\cite{star-3,star-4,star-5,star-6,star-7}. Scaling factors for each species are applied for visibility. Solid lines (blue online) are to guide the eye. Scaling systematic error of 10\% related to the trigger uncertainty is not shown.}
\label{fig:mt}
\end{figure}

A compilation of p+p data measured at $\sqrt{s}$=200~GeV is shown in Fig.~\ref{fig:mt}. Different sets of published and preliminary PHENIX data extracted from publications~\cite{ppg30,ppg55,ppg63,ppg64,ppg69,vic,max,nakamiya} are shown with different symbols. The STAR data~\cite{star-3,star-4,star-5,star-6,star-7} are shown with stars and crosses (red online). Figure~\ref{fig:mt} represents a unique and the most complete set of measurements from the RHIC experiments for 9 different identified mesons including LVM such $\rho$-, $\omega$-, K$^{*}$- and $\phi$-mesons. The measurements of the two experiments complete each other and generally are in good agreement within the systematic errors of the measurements shown in the plot. Note that the systematic errors related to the efficiency of the p+p trigger consisting about 10\% are not shown. At very low transverse momentum the results for the $\phi$-meson measurements from the STAR and PHENIX experiments are slightly different, however above 1~GeV/$c$ the agreement is much better. Also the published results for the $\rho$-meson invariant yield production measured by the STAR experiment is different from preliminary results of the $\omega$-meson spectra measured in the same $p_{T}$ range by the PHENIX experiment both in hadronic and leptonic decay channels.

\begin{figure}
\resizebox{0.49\textwidth}{!}{
\includegraphics{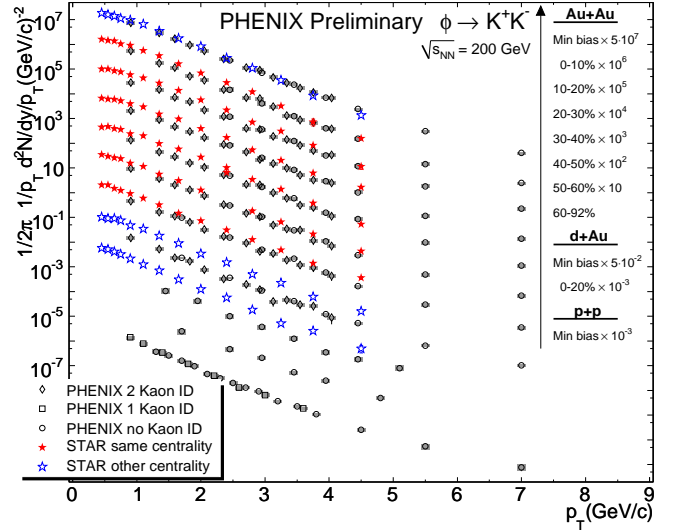}
}
\caption{The invariant $\phi$-meson spectra measured in different centrality classes in Au+Au collisions at $\sqrt{s_{NN}}=200$~GeV by the STAR collaboration are shown with star markers (color online). The invariant spectra measured by the PHENIX experiment are shown with a set of different markers (black online). Different markers represent three methods used by PHENI as explained in the text. Closed stars (red online) are the STAR results measured in the centrality bins coinciding with PHENIX, for open stars (blue online) the centrality bins are different. The data is taken from~\cite{max,ppg16,deb,star-8}}
\label{fig:phi} 
\end{figure}

The results of the measurement of the $\phi$-meson production at $\sqrt{s_{NN}}=$200~GeV in $^{197}$Au+$^{197}$Au collisions in the $K^{+}K^{-}$ decay channel carried out by the STAR and by PHENIX experiments are shown in Fig.~\ref{fig:phi}. The new preliminary PHENIX results substantially increase the $p_{T}$ range of the measurements up to 7~GeV/$c$ by not requiring particle identification for one or both K-meson. In the overlapping $p_{T}$ range the new measurements are consistent with the published PHENIX results~\cite{ppg16,deb} in which particle identification of both daughter particles was required. One should note that the new measurements done by PHENIX did not resolve the problem of the difference in the particle yields measured by STAR and PHENIX collaborations. One can see that for the same centrality bins the results of the two experiments differ from each other. The $\phi$-meson production rate measured by the STAR collaboration in Au+Au collisions is higher. It also results in a higher integrated $\phi$-meson yields since the slopes of the spectral lines agree. Unfortunately, up to now the inconsistency between the two experiments has not been resolved.

\subsection{Nuclear modification factor}
\begin{figure*}[ht]
\resizebox{0.98\textwidth}{!}{
\includegraphics{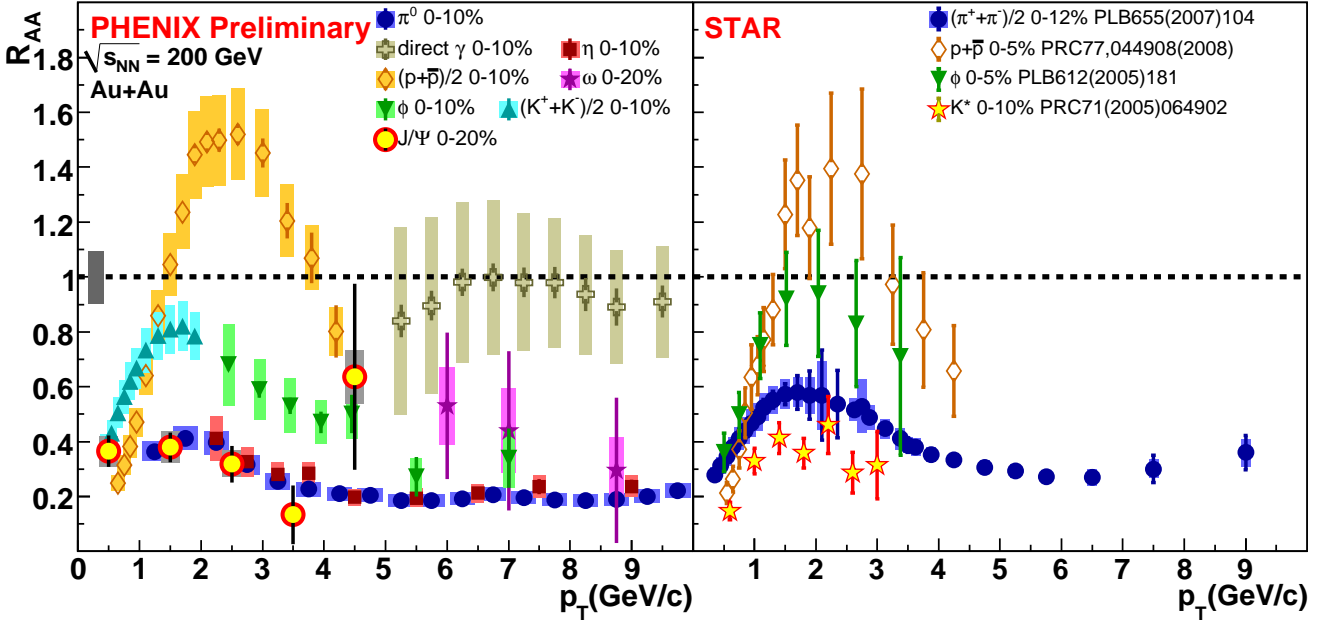}}
\caption{Nuclear modification factor $R_{AA}$ for $\gamma$, $\pi$, $\eta$ K, $\omega$, $K^{*}$ $\phi$, J/$\Psi$ measured in the most central Au+Au collisions at $\sqrt{s_{NN}}=$200~GeV. The results are from the PHENIX~\cite{vic,max,ppg68,ppg80,klaus} experiment  is in the left panel, and from the STAR~\cite{star-2,star-6,star-9,star-10} experiment is in the right panel. The experimental data for $\gamma$, $\pi$ and $\eta$ are shown in a shorter $p_{T}$ range than they are measured.}
\label{fig:raa} 
\end{figure*}

The Nuclear Modification Factor ($R_{AA}$) is defined as the ratio of the yields in A+A to p+p collisions scaled by the number of elementary binary collisions corresponding to a given A+A centrality class. An exact definition can be found elsewhere~\cite{raa}. The $R_{AA}$ is a parameter used to quantify the in-medium modification effects. The most complete set of identified particle $R_{AA}$ values in central Au+Au collisions at $\sqrt{s_{NN}}=$200~GeV measured by the STAR and the PHENIX experiments is shown in Fig.~\ref{fig:raa}. This figure presents the most recent published and preliminary results except for the strange baryons measured by STAR~\cite{star-1}.

The first results from RHIC~\cite{ppg3} experiments demonstrated that the $R_{AA}^{\pi}$ is significantly smaller than one, suggesting that at high-$p_{T}$ pions are subject to energy loss imposed by the media. Further measurements made at RHIC confirmed these results~\cite{star-9,ppg80}. It was later shown to be in drastic contrast to the $R_{AA}^{\gamma}$ which exhibit binary scaling out to very high $p_{T}$~\cite{klaus}. The conclusion derived from this comparison was that the high-$p_{T}$ suppression is not an initial state effect but is induced by the media. 

Another surprising result came out from the comparison of $\eta$ and $\pi$ suppression~\cite{ppg55,klaus}. Measured $R_{AA}$s of these two particles whose masses differ by a factor of four agree within errors in the entire range of the measurement. The result pointed out that the suppression mechanism is not coupled to the particle mass in the final state.

It was also shown that the $R_{AA}^{p+\bar{p}}$ of the proton~\cite{ppg30}, a baryon with a mass only twice larger than the mass of the $\eta$-meson has a very different suppression pattern and in fact a significant enhancement at $p_{T}$ of 2-3 GeV/$c$. The same effect holds for other heavier baryons~\cite{star-10} with even larger magnitude of the enhancement. This measurement suggested that the mechanism responsible for the suppression of high-$p_{T}$ particle production in Heavy Ion Collisions at RHIC energies is due to the number of quarks (mesons vs. baryons) and not due to the mass difference.

The best particle to test this hypothesis would be the $\phi$-meson with a mass similar to the mass of the proton. Both RHIC experiments measured the $R_{AA}^{\phi}$~\cite{star-6,max}. The measurements reveal that the magnitude of the suppression is different from both the lighter mesons and from the baryons. Discrepancies between the measurements of the two experiments mentioned earlier affect the $R_{AA}^{\phi}$ to a lesser extent because many systematic uncertainties cancel out. The high transverse momentum data from the PHENIX experiment seem to indicate that the $R_{AA}^{\phi}$ could reach the same level of suppression as measured for $\pi$- and for $\eta$- mesons, however it is not clear what causes the difference at lower $p_{T}$.

The answer to the question may come from the systematic investigation of the suppression patterns of other identified particles measured at RHIC. Preliminary data on the $\omega$-meson was presented by the PHENIX experiment~\cite{vic}. The $\omega$-meson has a mass smaller, but comparable to $\phi$, and has no strange quark content. Preliminary data on $R_{AA}^{\omega}$ exist only at very high $p_{T}$ and therefore do not help answering the question of the difference between the $\phi$-meson and other particles. The limited statistical significance of the ${\omega}$-measurement precludes a conclusive statement about an agreement between ${\phi}$ and ${\omega}$ Nuclear Modification Factors. The only information which can be extracted form the $R_{AA}^{\omega}$ is that its suppression magnitude does not disagree with that of the $\pi$ and $\eta$ $R_{AA}$s at $p_{T}$ above 5~GeV/$c$. A better measurement of $R_{AA}^{\omega}$ at lower $p_{T}$ would be very interesting to shed light on this puzzle.

Comparison between the $R_{AA}^{\phi}$ and the $R_{AA}^{K}$~\cite{ppg30} is informative to isolate the role of the strange quark in the suppression mechanism of the mesons. Currently such comparison is indirect since the measurements of K and $\phi$ do not cover a common $p_{T}$ range. Although one can continue the $R_{AA}$ of Kaons with the $R_{AA}^{\phi}$ such extrapolation can be only speculative. Extension of both nuclear modification factors to achieve overlap in the $p_{T}$ range does not confront any major detector limitations and therefore may be done in future.

Another surprising result comes from the measurement of the $R_{AA}$ for another vector meson, the $K^{*}$~\cite{star-5} published by the STAR experiment. The lifetime of $K^{*}$ is 10 times shorter than that of the $\phi$-meson. The $R_{AA}$ magnitude for this meson is twice lower compared to the STAR measurements for $\phi$ (as well as the PHENIX measurement for $R_{AA}^{K}$), although the mass of $\phi$ and $K^{*}$ differ by less than 15\% and all three particles have strange quarks. As pointed out in the STAR paper~\cite{star-5}, a smaller $R_{AA}^{K^{*}}$ can be due to re-scattering of the $K^{*}$ daughter products especially at low $p_{T}$.

The last mesonic nuclear modification factor measured at RHIC~\cite{ppg68} is the $R_{AA}$ of the J/$\Psi$-meson which is also plotted in Fig.~\ref{fig:raa}. At low $p_{T}$ the J/$\Psi$-meson shows the same magnitude of the suppression as the $K^{*}$-meson, but it is clear that drawing parallels between the suppression mechanisms of the charm particles and of the other mesons is not straightforward. Understanding of the $R_{AA}^{J/\Psi}$ is one of the most challenging goals of the RHIC program. It is possible that a better knowledge about the LVM $R_{AA}$s may help to resolve this puzzle.

%It is interesting to note that the only other $R_{AA}$ at low $p_{T}$ with the same magnitude of the suppression is that of the J/$\Psi$-meson measured by PHENIX collaboration~\cite{ppg68}. Understanding of the $R_{AA}^{J/\Psi}$ is one of the most challenging goals of the RHIC program. Better understanding of the LVM $R_{AA}$s may help the achieve this goal, but it is clear that drawing parallels between the suppression mechanisms of the charm particles and of the other mesons discussed in this section is not straightforward.

To summarize this section, new results recently presented for the measurement of the light vector mesons may shed light on the mechanisms of particle production in HIC. The inconsistencies in the measurements between RHIC experiments need to be resolved to allow better quantitative understanding of the data. Also an extension of the transverse momentum range measured for several particles, including $\phi$, K and protons would be very useful. Currently the existing results for $R_{AA}$ of different particles are not fully understood. The suppression patterns for different mesons and baryons allow for more than a single interpretation. A comparative analysis of existing data may help to better understand the physics of the Heavy Ion Collisions.


\end{document}